\documentstyle[amssymb,aps,12pt]{revtex}

\begin{document}
\draft
\author{O. B. Zaslavskii}
\address{The Erwin Schr\"odinger
International Institute for Mathematical Physics,\\ Boltzmanngasse 9,
A-1090, Wien, Austria\\
Permanent address:
Department of Mechanics and Mathematics, Kharkov V.N. Karazin's National\\
University, Svoboda Square 4, Kharkov 61077, Ukraine\\
E-mail: ozaslav@kharkov.ua}
\title{Boulware state and semiclassical thermodynamics of black holes in a cavity}
\maketitle

\begin{abstract}
A black hole, surrounded by a reflecting shell, acts as an effective
star-like object with respect to the outer region that leads to vacuum
polarization outside, where the quantum fields are in the Boulware state. We
find the quantum correction to the Hawking temperature, taking into account
this circumstance. It is proportional to the integral of the trace of the
total quantum stress-energy tensor over the whole space from the horizon to
infinity. For the shell, sufficiently close to the horizon, the leading term
comes from the boundary contribution of the Boulware state.
\end{abstract}



One of the brightest features of black holes is the fact that a
black hole possesses thermal properties such as the entropy and the
temperature. These entities acquire the literal meaning in the state of thermal
equilibrium (the Hartle-Hawking state). For the system to achieve this state,
two ingredients become essential. Firstly, one should take into account
the presence of quantum radiation around the hole. Quantum fields propagating in
a black hole background affect the geometry and, in particular, change the
surface gravity which determines the value of the Hawking temperature. Secondly,
Hawking radiation should be constrained inside a cavity that prevents quantum
fields from escaping to infinity. Thus, some overlap between quantum and boundary
effects should exist in black hole thermodynamics.
Consider a black hole enclosed inside a perfectly reflecting shell (microcanonical
boundary conditions). As the stress-energy tensor of quantum fields in this state
is bounded on the horizon, quantum backreaction leads to small corrections to
the geometry and Hawking temperature that can be calculated within the
perturbative approach. Such a program was realized for different types of
fields \cite{york85} - \cite{ds}. In doing so, it was usually implied that the region
outside the shell represents the usual vacuum, giving no contribution to
thermodynamics.

Meanwhile, actually the space outside the shell is not empty. A black hole
inside the shell, along with its Hawking radiation, acts as a source of the
gravitational field outside and curves spacetime. Therefore, vacuum
polarization is inevitable outside and represents the Boulware state (vacuum
with respect to the Schwarzschild time) rather than a pure classical vacuum.
In this state the average values $\left\langle T_{\mu }^{\nu }\right\rangle
\equiv T_{\mu }^{\nu }$ of quantum fields certainly contribute to the mass,
measured by a distant observer. Here the tensor $T_{\mu }^{\nu }$ is
supposed to be calculated in the main (one-loop) approximation as usual. The
correction to the mass due to the contribution of the vacuum polarization
outside the shell does not affect the geometry of the
spherically-symmetric configuration inside. The outer region does not
contribute to the entropy either since the Boulware state does not possess
thermal properties. However, by contrast with the mass and entropy, the
outer region should affect the Hawking temperature. Indeed, even for a pure
a classical Schwarzschild geometry a massive shell between the
horizon and infinity changes the Hawking temperature (it can be easily
seen if one matches the the metric inside and outside the shell). More than that, in
the situation under discussion the geometry deviates from the pure
Schwarzschild one due to backreaction of quantum fields. Thus, the value of
the Hawking temperature should feel the presence of the quantum fields in
the outer region. If the radius of the shell is large enough, the region
between the shell and infinity does not contribute to physical quantities
significantly, and neglecting vacuum polarization is quite reasonable
approximation. However, for a radius of the shell, compatible with the
horizon, the effect becomes essential.

The aim of the present paper is to find explicitly quantum corrections to
the Hawking temperature $T_{H}$, caused by these effects and, thus,
elucidate the influence of the Boulware state on black hole thermodynamics.

Consider the metric of a black hole:
\begin{equation}
ds^{2}=-U(r)dt^{2}+V^{-1}(r)dr^{2}+r^{2}(\sin ^{2}\theta d\phi ^{2}+d\theta
^{2})\text{.}  \label{1}
\end{equation}
From the Einstein equations it follows
\begin{eqnarray}
&&V(r)=1-\frac{2m(r)}{r}\text{, }m(r)=m+m_{q}(r)\text{,}  \label{2} \\
&&m_{q}(r)=4\pi \int_{2m}^{r}drr^{2}(-T_{0}^{0})\text{, }U=Ve^{2\psi }\text{,%
}  \nonumber \\
&&\psi =4\pi \int_{\infty }^{r}drr\frac{(T_{r}^{r}-T_{0}^{0})}{V(r)}\text{.}
\nonumber
\end{eqnarray}

It is assumed that a reflecting shell is placed at $r=R$, so the geometry
deviates from the Schwarzschildian one due to quantum corrections. For $%
r\rightarrow \infty $, $\psi \rightarrow 0 $ (provided the quantum
stresses decay rapidly enough) and the geometry approaches its
Schwarzschildian form. For $r<R$, where a black hole and thermal radiation
are present at the temperature $T_{H}$, the quantum fields are in the
Hartle-Hawking state. For $r>R$, the fields are in the Boulware state. As,
in general, on the boundary stresses, calculated in two different states, do
not coincide, there appears a jump. The total stress-energy $\theta _{\mu
}^{\nu }$, including that of the shell, reads

\begin{equation}
\theta _{\mu }^{\nu }=T_{\mu }^{\nu HH}\theta (R-r)+T_{\mu }^{\nu B}\theta
(r-R)+T_{\mu }^{\nu S}\text{, }  \label{t1}
\end{equation}
$\theta (r)$ is the step Heaviside function, $T_{\mu }^{\nu S}$ describes
the contribution from the shell. It is implied that we work in the
one-loop approximation, so the stresses (which are responsible for quantum
corrections) are calculated with respect to the unperturbed classical
background, i.e. the Schwarzschild geometry. Then it follows from the conservation law $%
\theta _{\mu }^{\nu };_{\nu }=0$ with $\mu =r$ in the Schwarzschild
background or from the general formalism \cite{ju} that nonzero components
of $T_{\mu }^{\nu S}$, necessary to maintain equilibrium, are equal to
\begin{equation}
T_{\phi }^{\phi S}=T_{\theta }^{\theta S}=-\frac{R}{2}\delta
(r-R)[T_{r}^{rHH}(R)-T_{r}^{rB}(R)]\text{.}  \label{t2}
\end{equation}

In general the Euclidean version of the metric (\ref{1}) possesses a conical
singularity at the horizon $r=r_{+}=2m$. The only way to avoid it is to set
the temperature equal to its Hawking value $T_{H}=(4\pi )^{-1}\kappa =(4\pi
)^{-1}[U^{\prime }(r_{+})V^{\prime }(r_{+})]^{1/2}$, where $\kappa \,$is the
surface gravity. Then
\begin{equation}
T_{H}=(8\pi m)^{-1}[1+8\pi r_{+}^{2}T_{0}^{0}(r_{+})]\exp [\psi (r_{+})]%
\text{.}  \label{3}
\end{equation}
Making use the $r$-component of the conservation law, one obtains in the
Schwarzschild background
\begin{equation}
\frac{1}{4\pi }\frac{\partial \psi }{\partial r}=r\frac{\theta
_{r}^{r}-\theta _{0}^{0}}{1-\frac{2m}{r}}=\frac{1}{m}[r^{2}\theta _{i}^{i}-%
\frac{\partial }{\partial r}(r^{3}\theta _{r}^{r})].  \label{4}
\end{equation}
Consider for definiteness the scalar massless field. Then explicit
approximate calculations of the stress-energy tensor show \cite{nf}, \cite
{acta} that in the Boulware state $T_{\mu }^{\nu }\thicksim r^{-6}$, when $%
r\rightarrow \infty $, so $r^{3}T_{\mu }^{\nu }\rightarrow 0$ as $%
r\rightarrow \infty $. Then it follows from (\ref{4}), (\ref{t1}), (\ref{t2}%
) that
\begin{equation}
\psi _{+}\equiv \psi (r_{+})=\frac{4\pi }{m}[\int_{\infty
}^{r_{+}}drr^{2}T_{i}^{i}-R^{3}[T_{\mu }^{\nu HH}(R)-T_{\mu }^{\nu
B}(R)]-r_{+}^{3}T_{r}^{rHH}(r_{+})]\text{.}  \label{p+}
\end{equation}
The back reaction strength is governed by the small parameter $\varepsilon =
\rlap{\protect\rule[1.1ex]{.325em}{.1ex}}h%
/m^{2}\ll 1$ which is assumed to cause small corrections to the
Schwarzschild metric: $\psi \ll 1$. Then, replacing $e^{\psi }$ by $1+\psi $
and taking into account the regularity condition $%
T_{0}^{0}(r_{+})=T_{r}^{r}(r_{+})$ which follows from (\ref{4}) and the
finiteness of $T_{\mu }^{\nu }$ at the horizon in the Hartle-Hawking state,
we obtain
\begin{eqnarray}
&&T_{H}=(8\pi m)^{-1}(1+\delta )\text{, }\delta =\delta _{1}+\delta _{2}
\label{5} \\
&&\delta _{1}=\frac{4\pi }{m}[R^{3}T_{r}^{rHH}(R)-%
\int_{2m}^{R}drr^{2}T_{i}^{iHH}]\text{,}  \nonumber
\end{eqnarray}
\begin{equation}
\delta _{2}=\frac{4\pi }{m}[-R^{3}T_{r}^{rB}(R)-\int_{R}^{\infty
}drr^{2}T_{i}^{iB}].
\end{equation}

It is convenient to rewrite (\ref{5}) in terms of the total mass
\begin{equation}
M_{tot}=m+m^{HH}(r^{+},R)+m^{B}(R,\infty )\text{,}  \label{tot}
\end{equation}
where
\begin{equation}
m^{HH}(r^{+},R)=-4\pi \int_{2m}^{R}drr^{2}T_{0}^{0HH}\text{, }m^{B}(R,\infty
)=-4\pi \int_{R}^{\infty }drr^{2}T_{0}^{0B}\text{.}  \label{mm}
\end{equation}
With the same accuracy,
\begin{eqnarray}
&&T_{H}=(8\pi M_{tot})^{-1}(1+\gamma )\text{, }  \label{6} \\
&&\gamma =\frac{4\pi }{M_{tot}}\{R^{3}[T_{r}^{HHr}(R)-T_{r}^{rB}(R)]-%
\int_{2m}^{\infty }drr^{2}T_{\mu }^{\mu }\}\text{,}  \nonumber
\end{eqnarray}
where we took into account that the conformal anomaly does not depend on the
state.

With (\ref{t1}), (\ref{t2}) at hand, it can be rewritten in terms of the
total stress-energy tensor as
\begin{equation}
\gamma =-\frac{4\pi }{M_{tot}}\int_{2m}^{\infty }drr^{2}\theta _{\mu }^{\mu }%
\text{.}  \label{gamma}
\end{equation}

This expression includes (i) bulk contributions from quantum fields in the
Hartle-Hawking state inside the shell, (ii) the boundary term, (iii) bulk
contributions from quantum fields in the Boulware state outside the shell.
In Ref. \cite{anom} only (i) and (ii) were calculated. Now we generalized
that result for the entire system to take vacuum polarization (iii) into
account.

The quantity $\gamma $ in (\ref{gamma})\ split to two parts - $\gamma _{1}$,
depending only on the $T_{\mu }^{\nu HH}$, and $\gamma _{2}$, depending only
on $T_{\mu }^{\nu B}$. The latter quantity has, for massless fields, the
structure \cite{nf}, \cite{acta}
\begin{equation}
T_{\mu }^{\nu B}=KT_{H}^{4}\left( \frac{r_{+}}{r}\right) ^{6}[\frac{A_{\mu
}^{\nu }}{\left( 1-\frac{r_{+}}{r}\right) ^{2}}+B_{\mu }^{\nu }]\text{,}
\label{ab}
\end{equation}
where tensors $A_{\mu }^{\nu }$ and $B_{\mu }^{\nu }$ are finite everywhere,
including the horizon and infinity, $K$ is the numerical factor, singled out
for convenience.

For large $R$ the term $T_{r}^{HHr}(R)$ tends to the constant, while $T_{\mu
}^{\mu }$ and $T_{\mu }^{\nu B}$ behave like $r^{-6}$, and $\gamma
_{2}/\gamma _{1}\thicksim \left( r_{+}/R\right) ^{6}$. Therefore, for $R\gg
r_{+}$ we return to the situation considered in \cite{york85} - \cite{ds}.
However, for $R\thicksim 2r_{+}$, the corrections due to the Boulware state
can be significant. When $R\rightarrow r_{+}$,
\begin{equation}
\delta \simeq -8\pi r_{+}^{2}T_{r}^{rB}(R)\simeq CT_{H}^{4}r_{+}^{2}(1-\frac{%
r_{+}}{R})^{-2}\text{, }C=-8\pi A_{1}^{1}(r_{+})K\text{.}
\end{equation}
For conformal fields $K=\frac{2\pi ^{2}}{45}$, $A_{1}^{1}(r_{+})=-\frac{1}{4}
$, $C=\frac{4\pi ^{3}}{45}$. The quantity $\delta $ can be rewritten as
\begin{equation}
\delta =\varepsilon D(1-\frac{r_{+}}{R})^{-2}\text{,}
\end{equation}
where the small parameter $\varepsilon \equiv \frac{%
\rlap{\protect\rule[1.1ex]{.325em}{.1ex}}h%
}{m^{2}}$ governs the backreaction strength and $D=C\frac{1}{4^{5}\pi ^{4}}=%
\frac{1}{11520\pi }.$ As $\delta $ is the product of the small and big
quantities, it is possible that $\delta \ll 1$ and, thus, perturbation
theory is still valid. Then the fractional contribution of the Boulware
state to the mass contains an additional factor $(1-\frac{r_{+}}{R})$ and is
also small, so with the same accuracy $\gamma \simeq \delta $. We can see
that, for shells, sufficiently close to the horizon, the Boulware
contribution dominates the correction. In doing so, $\gamma >0$. If the
radius of the shell decreases further, $T_{\mu }^{\nu B}$ diverges like $(1-%
\frac{r_{+}}{R})^{-2}$, quantum backreaction fails to be small, and the
perturbation theory ceases to work.

There are two points, typical of treatment in an infinite space. Firstly,
usually the Hartle-Hawking and Boulware states appear in essentially different
contexts: the first one is relevant for a black hole metric, while the second one
applies to the background, typical of a relativistic star. Secondly, the
relevancy of vacuum polarization in black hole thermodynamics implies, as a
rule, the presence of massive fields. We saw, however, that account for the
finiteness of the system, containing a black hole, leads to the overlap
between both types of states, so the Boulware state does affect black hole
thermodynamics even in the case of massless fields.

I thank for hospitality Erwin Schr\"odinger International Institute for Mathematical Physics,
where this work has been brought to completion.

\end{document}